\input{aipcheck}
\documentclass[
,final            
  ]
  {aipproc}
\layoutstyle{6x9}
\usepackage{amsmath,amssymb,amsfonts}

\def\IM{\mathop{\Im{\rm m}}\nolimits}
\begin{document}
\title{Can one improve the Froissart bound?
\footnote{Dedicated to the memory of L.\ Lukaszuk, G.\ Sommer,
and F.J.\ Yndurain}
}
\author{\textbf{Andr\'e Martin}}{address={
CERN, Theory Division, CH1211 Gen\`eve 23, Switzerland}}
\date{\small Talk given at Diffraction 2008, Lalonde-les-Maures, September 2008}

\begin{abstract}
We explain why we hope that the Froissart bound can be improved, 
either \emph{qualitatively} or, more likely, \emph{quantitatively}, 
by making a better use of unitarity, in particular \emph{elastic}
unitarity. In other instances (Gribov's theorem) elastic unitarity played 
a crucial role. A preliminary requirement for this is to work with an 
appropriate average of the cross-section, to make the problem well 
defined. This is possible, without destroying the Lukaszuk--Martin bound.
\end{abstract}
\maketitle

Many of my friends complain that, in the
Froissart bound \cite{1,1bis} on total cross-sections, the factor in front of the logarithm square is
much too big. One of them is Peter Landshoff \cite{2} present at
this conference. Indeed the bound that we obtained with the late Lesek
Lukaszuk \cite{3}
\begin{equation}
\sigma_\text{total}< {\pi\over m_\pi^2}\,\left(\ln s\right)^2~,
\end{equation}
is perhaps  500 times larger than what is
found in fits of the proton{}-(anti) proton experimental
cross{}-sections \cite{4}.
People
should remember that this is only a bound in which the mass of the pion
appears because the two{}-pion threshold is the first one appearing in
the $t$-channel. 
However, we believe that there is some hope for
improvement. ``We'' means Joachim Kupsch, Jean{}-Marc Richard, Shasanka
Roy and me. The reason is that so far there is no example of amplitude
which has the right analyticity properties in all channels and
satisfies ``unitarity'', which means for the partial wave
amplitudes
\begin{align}
\IM{f_\ell}&\ge  |f_\ell|^2\quad \text{for all energies}~,\\
\IM{f_\ell}&= |f_\ell|^2\quad \text{in the elastic region}~,
\\
\intertext{with}
F(s, \cos \theta)&={\sqrt{s}\over2k}\,  \sum_\ell (2\ell+1)\, P_\ell( \cos\theta )  \,f_\ell(s)~,\\
A(s, \cos \theta)&={\sqrt{s}\over2k}\,  \sum_\ell (2\ell+1)\, P_\ell( \cos\theta )\IM{f_\ell(s)}  ~,
\end{align}
where $s$ is the square of the c.m.\  energy, $k$ the c.m.\
momentum, and $\theta$ the c.m.\ scattering angle.

We realize that (2) and (3) represent only a small fraction of
unitarity, but these are the only ingredients we can use if we don't
not want to involve many{}-body amplitudes, much too complicated to
handle.

Then, using positivity one can prove that $F$ is
analytic in $t$ \cite{5} for 
\begin{equation}
t<4\,m_\pi^2~.
\end{equation}
The maximum value of $t$ in (6) is given by
the prescription of Sommer, or the more refined approach of Bessis and
Glaser \cite{6}, $t$ being given
\begin{equation}
t= 2\,k^2\left(\cos\theta-1\right)~.
\end{equation}

One can also prove  that if $F$ is bounded by $s^N$, $N$ even, for $t=0$ it has also the same bound for $t=4\,m_\pi^2$.
Then, using unitarity, one gets the Froissart bound with a constant
depending on $N$. The Froissart bound means that $N=2$ for $t = 0$. Then,
using the fact that the number of subtractions does not change for
positive $t,$ one can repeat the cycle and end up with the bound (1).

The question is whether this bound
can be improved, i.e., replacing the exponent of the logarithm by an exponent smaller
than 2, or \textit{quantitatively}, i.e., replacing the factor in (1), that Peter
dislikes, by something much smaller. 
 Kupsch \cite{7} has constructed an
example satisfying  (2) but not
(3), in which the bound  (1) is anyway certainly not saturated
because, to simplify things,  he takes the partial wave amplitudes
to be less than 1/2 instead of 1. It is not even clear that (1) could
be saturated using only the ``inelastic'' unitarity  (2). However, the
greatest hope for improvement comes from the fact that nobody has been
able construct a model saturating the Froissart bound and satisfying
both (2) (``inelastic unitarity'') and (3) (``elastic unitarity'').
Atkinson \cite{8} has
succeeded to construct a model satisfying (2) and (3), but his
cross{}-section behaves like
\begin{equation}
[\ln s]^{-3}~.
\end{equation}
The importance of ``elastic'' unitarity is
very well illustrated by Gribov's theorem \cite{9}
that
contrary to the general belief before 1960, it is impossible to have a
scattering amplitude such that
\begin{equation}
F(s,t)\sim s\,f(t)~,\quad
\text{with }  f(0)  \text{ non\  zero}~,
\end{equation}
or, in words, such that the cross{}-section
tends to a non zero constant with a finite width of the diffraction
peak at infinite energy. His proof rests very heavily on ``elastic''
unitarity and Jean{}-Marc Richard and I 
\cite{10} and, also,
Joachim Kupsch \cite{11}, have
constructed examples in which only (2) is required, and where the
scattering amplitude does behave like $s\,f(t)$  at high energy. Our
example is
\begin{equation}
F=\text{Const}\left[ F_s+F_t + F_u\right]~,
\end{equation}
with 
\begin{equation}
F_s=\left[4-\sqrt{(4-t)(4-u)}\right]  \exp\left[-2(4-s)^{1/4}\right]~,
\end{equation}
where $m_\pi=1$.

My belief is that the Froissart bound cannot be improved. May be that I am not objective because
this would kill the model of my friends Claude Bourrely, Jacques
Soffer, and Tai Tsun Wu\cite{Bourrely:2002wr}. I would
like to present here a sketch of an idea, which might or might not
work, to construct an amplitude which saturates the Froissart bound and
satisfies (2) and (3). We take the case of pion{}-pion scattering.  We assume the validity of Mandelstam representation
\begin{equation}
F(s,t)=\int\!\!\int \mathrm{d} x\,\mathrm{d}y \, {\rho(x,y)\over (x-s)(y-t)}
+ \text{circular permutations}+ \text{substractions}~,
\end{equation}
with $\rho(s,t)=0$ for $s<16$ and $t<16$.

 The amplitude is \textit{real}  for  $s <16$, and we impose condition
(2) \emph{only} $s>16$.  I believe that the construction of
Kupsch can be carried out completely in this framework, though I tried to
have his opinion on that without success (he is trying another
approach!). Anyway, suppose it works, and suppose $F$ saturates
qualitatively the Froissart bound. Then, if $F$ is a solution,
$\lambda\,F$ is also a solution, for
$\lambda<1$. One can now try to correct for the fact that unitarity
has not been taken into account in the elastic strip by using (3) (with $f_\ell$ real on the right hand side ), and obtain a first approximation
of the absorptive part in the elastic region, which will be of order
$\lambda^2$.
Then one can recalculate a new amplitude using dispersion
relations. This new  amplitude differs  from the previous one by
something of the order
$\lambda^2$.
 One can hope that for
$\lambda$
sufficiently small a convergent iterative procedure is possible. There
are difficulties in the first steps, but they might be overcome.

If this works, this means that there is no
\textit{qualitative} improvement possible. This would be good for our friends,
Bourrely--Soffer--Wu. However, there is room for a
\textit{quantitative} improvement. We do not know, at the
present time,  how to proceed to do this. However a preliminary
requirement is to work with an
\textit{average}
of the cross{}-section. Among the experts of the subject, everybody
knows that the Froissart  bound is not
\textit{local}. The Froissart bound follows from the
fact that the integral of $A(s, t=4)$ divided by $s^3$ is
\textit{convergent}. In fact it is possible to construct
examples in which, for a given, arbitrarily high, energy the total
cross{}-section is infinite locally. The Froissart bound applies to an
average of the cross{}-section over an energy interval which can be
very narrow ,  an arbitrary negative power of the energy. \ However
the constant in the bound depends on the choice of the interval. All
this was realized long ago by Common and Yndurain \cite{12}. The
constant in (1) holds for a sequence of energies with asymptotic
density unity if $A(s, t=4)$ is continuous. What we have found is that if
we want to avoid this assumption, and if we want this average to
satisfy the bound  (1) the interval over which we take the average
must be of the order of $s$,  for instance $s$ to $2\,s$. The benefit of
using this average is that the scale factor in the logarithm is fixed
as a function of \textit{low-energy} parameters in the $t$ channel.
Therefore, another criticism to the Froissart bound is eliminated. In
fact, since this talk was given, I realized that an interval of the
order of $s/\sqrt{\ln s}$, is also acceptable, which means that the interval
becomes much smaller than $s$ at high energies.

Now what to do next? Honestly, we don't know. S.M.\  Roy
would like to try a variational approach,  assuming Mandelstam
representation, and varying  the completely inelastic part of the
double spectral function, i.e., the double spectral function minus what
is obtained from elastic unitarity in the $s$ and $t$ channels. I apologize
for not being more precise! This is only a kind of progress report.

\subsubsection*{Acknowledgments} The work presented here was carried with the support of 
the CEFIPRA (IFCPAR),  project No:  3404-3 (Indo-French collaboration). 
Hospitality at the I.H.E.S. ( Institut des Hautes Etudes Scientifiques, 
Bures sur Yvette) is also acknowledged.

\end{document}